# Wireless Application Protocol Architecture Overview

Seyed Hossein Ahmadpanah

Department of Computer and Information Technology, Mahdishahr Branch, Islamic Azad University, Mahdishahr, Iran.

Abdullah Jafari Chashmi

Department of Electrical and Telecommunications Engineering Technology, Mahdishahr Branch, Islamic Azad University, Mahdishahr, Iran.

Vahid Jahan

Department of Computer and Information Technology, Mahdishahr Branch, Islamic Azad University, Mahdishahr, Iran

*Abstract*— **WAP protocol is a set of communication protocols application environment and wireless devices. WAP model more than a WAP gateway. Suitable for wireless communication environment.**

## I. INTRODUCTION

Wireless Application Protocol (WAP) is through continuous efforts to get into as a result, it provides an industry specification, developed to apply to a variety of wireless communication networks and business applications. [1]

WAP provides for a variety of network protocols and application frameworks wireless devices, these devices including mobile phones, pagers machine, a personal digital assistant (PDA) and the like. This specification not only extends the mobile networking technologies (such as digital data networking standards) and Internet technologies (such as XML, URL, scripts, and various content formats), but also to promote their development.

WAP direction for the Forum is to enable operators, manufacturers and service content developers to quickly and flexibly build their own advanced, differentiated services.

WAP goal of the Forum is:

- Provide digital cellular phones and other wireless terminals Internet content and advanced data services.
- Develop a global wireless protocol specification that works on a variety of wireless network technology.
- Able in a wide range (including multiple bearer networks and device types) to generate content and applications.
- where needed, the integration and expansion of existing standard variety of applications and technologies.

WAP architecture specification is designed to substantially meet the proposed system architecture and protocol WAP objectives of the Forum, it may WAP to start understanding and a series of technical specifications thus generated acts. In reference to the corresponding specification on the basis of, WAP system architecture specification also provides an overview of the different technologies, and thorough preparation for further research.

## II. WAP ARCHITECTURE

1. WAP communication model overview

Traditional WWW client / server (C / S) structure. Client Web browser sends a service request to the network server, using standard data model to respond. Compared to traditional communication and WWW, WAP also uses a client / server mode.[2] But the biggest difference is between the two: between the client and server, WAP model more than a WAP gateway. By then the client communicates with the server resources (Origin server) WAP gateway. At the same time, information between the client and the WAP gateway delivery is also different from the traditional way of information between client and server exchange.

WAP content and applications using the WWW similar pattern definition content transmission also uses a set of communication protocols and WWW similar standard communication protocols. Similar micro-browser and standard mobile terminal WEB browser, is responsible for coordinating the interface with the user. Taking into account the bandwidth limitations of wireless networks, the need for information between the client user agent with WAP gateway delivery (both request and response) coding, in order to reduce network traffic, maximize the use of the wireless network slow data transfer rate.

WAP gateway is a WAP proxy. WAP wireless technology to connect using a proxy domain and the WWW. WAP proxy typically consists of two main functions:

1. Protocol conversion - in charge of the WAP protocol stack (WSP, WTP, WTLS, and WDP) the request into a WWW protocol stack (HTTP and TCP / IP) requests.

2. Content encoding and decoding - is responsible for the content encoder converts compression encoding

formats WAP content, thereby reducing the amount of data transmitted over wireless networks.

Standard models include WAP client, WAP and WAP proxy server. But WAP architecture can support other configurations. For example, the WAP proxy functionality is included in WAP server, so that you can achieve the client and server security Secretary-end connectivity. [3]

2. WAP architecture Composition

WAP application architecture for the development of mobile communication equipment provides a scalable, extensible environment. It uses similar to the TCP / IP protocol stack hierarchical design, but modified and optimized to suit a wireless communication environment. [4] Wherein each layer protocols define a standard interface that can be called the upper layer protocol may also be other services and applications direct access.

The following are the layers WAP architecture brief introduction

1. WAE: Wireless Application Environment

WAE application development environment is a common sense, easy and efficient support to develop and run applications on different wireless communication network. A typical WAP application system includes three entities: a mobile terminal having a user agent functionality, protocol conversion of WAP proxy (Proxy) to provide application services and the source server (origin server).

2. WSP: Wireless Session Protocol

WSP WAE using a uniform interface to the application layer provides two types of services: connection-oriented service based on WTP and WDP-based connectionless service. Currently, WSP comprise suitable browser application service (WSP / B).

WSP / B provides features include:

- compressed coded representation of the HTTP l. Request semantics;
- long session state 3
- Session suspend and resume negotiation and protocol functions.

WSP / B allows for connection WAP standard HTTP client and server via the WAP proxy.

3. WTP: Wireless Transaction Protocol

WTP provides a lightweight, transaction-oriented services. WTP provides the following features effectively on a secure or non-secure wireless datagram networks:

- three types of transaction services, including: unreliable one-way request, reliable way request and reliable two-way request - A celebration affair;
- (Optional) user-to-user reliability, i.e. the user for each piece of information received is confirmation;
- (Optional) band data response;
- PDU (Protocol Data Unit) cascaded and delayed responses;
- asynchronous transactions.

4. WTLS: Wireless Transport Layer Security protocol

WAP architecture is noteworthy that adds a layer of security. It builds on TCP / IP architecture no security mechanism and thus bring great threat to network communication lessons, specially set up a security layer to protect the security of communications. [5]

WTLS is based on Transport Layer Security (TLS) security protocol. WTLS is optimized for narrow bandwidth wireless communications and provide secure transport services up in WDP basis. The main function of WTLS provides:

- Data integrity: WTLS to ensure that data transmission is not modified and destruction between the mobile terminal and application server;
- privacy: WTLS ensures that data between the mobile terminal and application server transmission is private and can not be understood by any third party received the data;
- authentication: WTLS ensure the authentication of the mobile terminal and the server;
- Denial of Service protection: WTLS contains a set of tools that can detect and reject duplicate transmission or unsuccessful verification of data, so that many typical denial of service attacks are more difficult to achieve effective protection of the upper-layer protocols.

Applications can be selectively enabled or disabled WTLS features depending on the characteristics of their own security requirements and the underlying network.

5. WDP: Wireless Datagram Protocol

WAP architecture as the transport layer protocol, WDP use the underlying network carrier to provide a consistent and transparent data transfer service for upper-layer protocols. WDP upper layer protocol to shield the details of the underlying network, so that the upper layer protocols can be used independent of the underlying network and the way work, but also to the upper application can be ported between different network platforms.

6. BEARER: the underlying bearer network

WAP protocol was originally designed is to be able to run on various existing carrier services, such as: short message service (SMS), circuit switched data (CSD) and the like. The underlying bearer network provides up different throughput, bit error rate and delay service, these differences are due to the presence of WDP layer and upper layer protocol transparency. WDP specification and supports bearer network technology allows WAP protocol running on each of the carrier used has been described. Of course, WDP will be supported by the carrier with the advent of new technologies and constantly changing over time.

7. other services and applications:

WAP layered architecture enables other services and applications using the WAP good interface protocol stack through a set of function definitions. External applications can directly access the protocol stack in the session layer, transaction layer, security and transport layers. Such direct call service provided by the layers, which greatly facilitates the development of a variety of applications.

### III. CONCLUSION

Looking back on the Internet in the past 10 years to promote overall economic development, and now the Internet is becoming increasingly popular, whether this can be seen by the WAP's future? Although WAP industry experienced many ups and downs, is still in its infancy, there are a lot of people deeply concerned about the future of the WAP, but I believe that with the development of mobile technology, WAP services targeted creative operators to give appropriate support, and worked out a reasonable operating and profit model, WAP will be the same as it is now the Internet to create brilliant.